\documentclass[%
reprint,
nofootinbib,
 amsmath,
 amssymb,
 aps,
]{revtex4-2}

\usepackage{graphicx}
\usepackage{dcolumn}
\usepackage{bm}

\usepackage{txfonts}
\usepackage{xspace}
\usepackage{hyperref}

\newcommand{\beq}{\begin{equation}}
\newcommand{\eeq}{\end{equation}}
\newcommand{\beqa}{\begin{eqnarray}}
\newcommand{\eeqa}{\end{eqnarray}}
\newcommand{\mpc}{$h^{-1} \mathrm{Mpc}$}
\newcommand{\mpcden}{$h^{3} \mathrm{Mpc}^{-3}$}
\newcommand{\msun}{$h^{-1} {\rm M}_{\odot}$}

\newcommand{\eg}{e.g.,\xspace}

\def\der{{\rm d}}
\usepackage[normalem]{ulem}
\usepackage{svg}
\usepackage{comment}

\begin{document}


\title{Velocity recostruction with graph neural networks}

\author{Hideki Tanimura}
 \email{hideki.tanimura@ipmu.jp}
\affiliation{%
 Center for Data-Driven Discovery, Kavli IPMU (WPI), UTIAS, The University of Tokyo, Kashiwa, Chiba 277-8583, Japan
}%


\author{Albert Bonnefous}
\affiliation{
 \'{E}cole Normale Sup\'{e}rieure de Lyon, 15 parvis Ren\'{e}-Descartes 69342 Lyon Cedex 07, France
}%
\author{Jia Liu}
\affiliation{%
 Center for Data-Driven Discovery, Kavli IPMU (WPI), UTIAS, The University of Tokyo, Kashiwa, Chiba 277-8583, Japan
}%

\author{Sanmay Ganguly}
\affiliation{%
 ICEPP, The University of Tokyo. Hongo, Bunkyo-ku, Tokyo 113-0033, Japan
}%
\affiliation{%
 Indian Institute of Technology Kanpur, Kalyanpur, Kanpur 208016, Uttar Pradesh, India
}%

\date{\today}

\begin{abstract}
In this work, we seek to improve the velocity reconstruction of clusters by using Graph Neural Networks---a type of deep neural network designed to analyze sparse, unstructured data. In comparison to the Convolutional Neural Network (CNN) which is built for structured data such as regular grids, GNN is particularly suitable for analyzing galaxy catalogs. In our GNNs, galaxies as represented as nodes that are connected with edges. The galaxy positions and properties---stellar mass, star formation rate, and total number of galaxies within 100~\mpc---are combined to predict the line-of-sight velocity of the clusters. To train our networks, we use mock SDSS galaxies and clusters constructed from the Magneticum hydrodynamic simulations. Our GNNs reach a precision in reconstructed line-of-sight velocity of $\Delta\varv$=163 km/s, outperforming by $\approx$10\% the perturbation theory~($\Delta\varv$=181 km/s) or the CNN~($\Delta\varv$=179 km/s). The stellar mass provides additional information, improving the precision by $\approx$6\% beyond the position-only GNN, while other properties add little information.  Our GNNs remain capable of reconstructing the velocity field when redshift-space distortion is included,  with $\Delta \varv$=210 km/s which is again 10\% better than CNN with RSD. Finally, we find that even with an impressive, nearly 70\% increase in galaxy number density from SDSS to DESI, our GNNs only show an underwhelming 2\% improvement, in line with previous works using other methods. Our work demonstrates that, while the efficiency in velocity reconstruction may have plateaued already at SDSS number density, further improvements are still hopeful with new reconstruction models such as the GNNs studied here. 
\end{abstract}

\maketitle


%

\section{Introduction}
\label{sec:intro}

The velocity field of the large-scale structure is a useful probe of cosmological and astrophysical parameters. Reconstructing the velocity field was an essential step in previous works to e.g. sharpen the baryon acoustic oscillations (BAO) peaks \citep{Blazek2016, Schmidt2016, Slepian2018} and to extract the kinematic Sunyaev-Zel’dovich (kSZ) \citep{Sunyaev1980} via cross-correlations with the Cosmic Microwave Background (CMB) measurements \citep{Cooray2001, DeDeo2005, Shao2011, Schaan2016, Schaan2021, Planck2016IRXXXVII, Munchmeyer2019, Lim2020, Nguyen2020, Tanimura2021, Tanimura2022, Bolliet2023, Cayuso2023}. The BAO signal originates from the acoustic density waves in the primordial plasma. It is used as a standard ruler to measure the expansion history of the universe \citep{Percival2007, Percival2010, Beutler2011, Blake2011, Anderson2012, Anderson2014, Dawson2013, Ross2015, Alam2017, Abbott2018, Abbott2019, Ata2018, Bautista2018, Bautista2021, Agathe2019, Blomqvist2019, Bourboux2020, Gil2020, Hou2021}. The kSZ signal originates from the scattering of CMB photons off coherently moving electrons. Because the kSZ effect is a particularly powerful probe of the electron distribution in under-dense regions, it was used to investigate the ``missing baryon problem'' \citep{Hill2016,Ferraro2016,Munshi2016, Lim2020} and to constrain baryonic feedback \citep{Park2018, Kuruvilla2020, Tanimura2022, Amodeo2021, Schaan2021, Lee2022}. 

In the past, the velocity field was typically reconstructed by solving the linearized continuity equation on the galaxy density field \citep{Zaroubi1995, Zaroubi1996, Zaroubi1999,  Branchini1999, Wang2012, Tanimura2021}. While the method works well on large scales where density fluctuation is small (``linear''), three issues complicate the procedure on small scales ($\lesssim$~30~\mpc). First, structure formation becomes nonlinear on these scales and hence is not fully captured in the linearized continuity equation \citep{Kitaura2012a, Kitaura2012b}. Second, to obtain the underlying matter field, we need to correct the galaxy bias, which becomes nonlinear on small scales and often requires numerical simulations to calibrate \citep{Benson2000, Berlind2002, Weinberg2004, Nuza2013, Kitaura2016, Rodriguez2016}. Finally, the line-of-sight (LOS) positions of the galaxies are measured in redshift, and hence we need to correct for redshift-space distortion (RSD) to obtain the comoving distance~\citep{Jackson1972, Kaiser1987}. 

In this work, we use the Graph Neural Network (GNN)---a deep neural network particularly suitable to sparse, heterogeneous data---to reconstruct the velocity field. The application of machine learning to velocity reconstruction is not new. Previously, \cite{Hong2021, Wu2021, Tanimura2022, Wu2023} performed such tasks using the Convolutional Neural Network (CNN). However, there are several limitations with CNNs in handling galaxy catalogs that motivated us to experiment with the GNN architecture \citep{Zhou2018}. First, CNNs are optimized for grid-like data but less so for irregularly structured data such as galaxy catalogs. Second, CNNs focus mainly on local features and hence likely to overlook global patterns. Third, CNNs are trained on a single data type, i.e. galaxy positions, and hence unable to incorporate other potentially useful galaxy features such as color, shape, stellar mass, and star formation rate. In comparison, GNNs can handle unstructured data of arbitrary size and complex topology,  capture global patterns, and allow us to incorporate multiple data types such as galaxy positions and properties. Thanks to these advantages, GNNs with galaxies have been recently used to predict halo masses \citep{Pablo2022} and the cosmic matter density $\Omega_m$ \citep{Santi2023, Shao2023}.  In this work, to quantify the performance of GNNs on realistic observations, we reconstruct the velocity fields using GNNs trained on mock galaxies of the Sloan Digital Sky Survey (SDSS) and the Dark Energy Spectroscopic Instrument (DESI), and compare them with the true velocity fields. 

This paper is organized as follows. Section \ref{sec:method} describes the mock simulations used in this study, the general GNN architecture, the specific network design for our galaxy catalogs, and our training procedure. The results are presented in Section \ref{sec:results}. We conclude in Section \ref{sec:conclusion}.


\section{Methodology}
\label{sec:method}

In this section, we describe the SDSS and DESI \citep{DESI2016} mock catalogs used in this work, the basics of GNN, the specific GNN architecture we adopted to model the galaxy distribution, and our training procedure.

\subsection{SDSS and DESI mocks from the Magneticum simulations}
\label{sec:magneticum}

We construct SDSS and DESI mock galaxy catalogs from the Magneticum simulations\footnote{\protect\url{http://www.magneticum.org/simulations.html}}, which are among the largest cosmological hydrodynamic simulations~\citep{Hirschmann2014, Dolag2015}. 
They are based on the standard $\Lambda$CDM cosmology with \{$\Omega_{\rm m}$, $\Omega_{\rm b}$, $h$\}=\{0.272, 0.046, 0.704\}~\cite{Komatsu2011}. Several simulation boxes with different sizes and resolutions are available. 
In this work, we use their \texttt{Box2b} with a box size of 640~\mpc\ and a total number of particles $N_p=2 \times 2880^3$, with equal numbers of dark matter and baryon particles. The simulations include post-processed data of galaxy and cluster catalogs. 

To obtain realistic SDSS-like galaxies, we removed clusters with $M_{500} < 10^{13}$~\msun\ and galaxies with $M_{*} < 2.0 \times 10^{11}$~\msun\ from the main catalogs, leaving us with 47,132 clusters and 77,101 galaxies.
In addition, to investigate the effect of varying galaxy number density, we create DESI-like mock catalogs by removing galaxies with $M_{*} < 1.4 \times 10^{11}$~\msun, leaving 133,721 galaxies in our sample. 
Throughout this paper, we focus on simulations at $z\approx$ 0.47\footnote{``snapshot 26'' for  \texttt{Box2b}.} to compare our results to those using CNN as done in ~\cite{Tanimura2022}~(hereafter T22, for more details, see Sec.~3 in T22). 

\subsection{Graph neural network}
\label{sec:gnn}

A galaxy catalog can be naturally represented by a graph, where galaxies as \textit{nodes} are connected by \textit{edges}. Several \textit{features} can be added as node attributes, \eg 3-dimensional (3D) position, stellar mass, and star formation rate. An example graph centered at a galaxy cluster is shown in Fig.~\ref{fig:graph}. 

With a set of $N$ nodes, a graph can be described with an $N \times N$ adjacency matrix $G_{ij}$, where $G_{ij}=1$ if nodes $i,j$=$1,2,...N$ are connected by an edge and $G_{ij}=0$ otherwise. 
The neighborhood of a node $i$ is defined as 
\begin{equation}
    \mathcal{N}_i=\{j\:|\:G_{ij}=1 \}. 
\end{equation}

Each node receives information from its neighborhood via the message-passing scheme. In this scheme, the neighborhood's features (``messages'') are passed through a multilayer perceptron (MLP)---a fully connected neural network---and aggregated to form hidden feature vectors as 
\begin{equation}
    \mathbf{h}_i=\bigoplus_{j\in \mathcal{N}_i}\psi(\mathbf{x}_i,\mathbf{x}_j), 
    \label{eq:mp}
\end{equation}
where $\bigoplus$ is the differentiable, permutation invariant aggregation function (\eg the maximum, the mean, or the sum), $\psi$ is the message-passing MLP, and $\mathbf{x}_i$ and $\mathbf{x}_j$ are the feature vectors of the node itself and its neighbor $j$, respectively. 

To obtain the output quantities $\mathbf{f}(\mathcal{G})$---cluster velocities in our case---the hidden feature vectors $\mathbf{h}_i$ and the global feature vector $\mathbf{u}$ are passed through a second MLP $\phi$,
\begin{equation}
    \mathbf{f}(\mathcal{G})=\phi\left(\bigoplus_{i\in\mathcal{G}}\mathbf{h}_i\, ,\mathbf{u}\right), 
    \label{eq:gp}
\end{equation}
where $\mathbf{u}$ describes global features of the graph such as the total number of nodes and that of edges. 

Finally, the network is trained to minimize the loss between the output (\eg the peculiar velocity of a cluster) and its true value.

\subsection{Galaxy distribution as a graph}
\label{sec:mygnn}

In this work, we focus on predicting the LOS velocities of clusters, instead of the full 3D velocities. Galaxy clusters are the largest gravitationally bound structures in the Universe. 
The LOS velocity is particularly relevant to the kSZ observations for upcoming high-resolution CMB experiments such as the Simons Observatory~\citep{Ade2019} and CMB-S4 \citep{Abazajian2019} to constrain cosmology and astrophysics. The kSZ signal is sensitive to the LOS electron momentum, 
\beq
\frac{\Delta{T}_{\rm kSZ}}{T_{\rm CMB}} = - \sigma_{\rm T} \int n_{\rm e} \, \left( \frac{\bm{\varv}\cdot \bm{\hat{n}}}{c}  \right) \, \der{l} 
\approx - \tau \left( \frac{\bm{\varv} \cdot \bm{\hat{n}}}{c} \right),
\label{eq:ksz}
\eeq
where $\sigma_{\rm T}$ is the Thomson scattering cross section, $c$ is the speed of light, $n_{\rm e}$ is the electron number density, and $\bm{\varv} \cdot \bm{\hat{n}}$ is the peculiar velocity of electrons along the LOS. The integral $\tau = \sigma_{\rm T} \int n_{\rm e} \der{l} $ can be performed given that the typical correlation length of electron velocity (given by $\bm{\varv} \cdot \bm{\hat{n}}$), 80–100~\mpc~\cite{Planck2016IRXXXVII}, is much larger than that of the gas density~($\approx$5~\mpc). Therefore, the reconstructed LOS velocity field can help break its degeneracy with the electron density in kSZ observations. 

To construct the graph for each cluster, we define the nodes to be galaxies within 100~\mpc\ of the cluster. The scale was chosen to match the typical correlation length of peculiar velocities of 80--100~\mpc~\citep{Planck2016IRXXXVII}. Galaxy pairs within 15~\mpc\ are connected by edges\footnote{While theoretically, all galaxies can be fully connected in the graph without the 15~\mpc\ limit, realistically the graph quickly becomes too large to store in the GPU memory.}. 

\begin{figure*}
\centering
\includegraphics[width=0.8\linewidth]{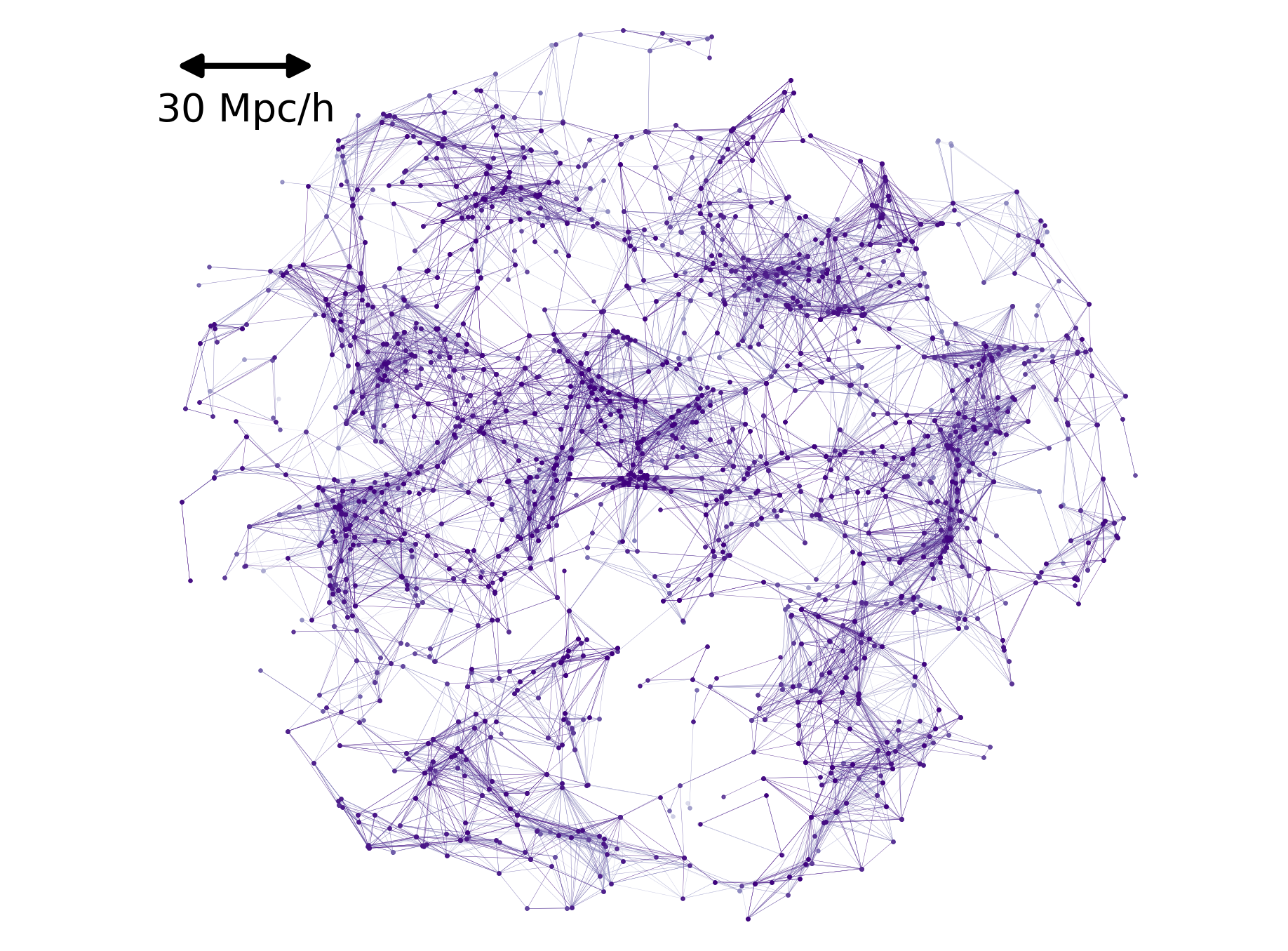}
\caption{An example graph centered at one cluster, projected onto a 2D plane. Galaxies within  100 $h^{-1}$Mpc of the cluster are represented as nodes. Galaxy pairs within 15~\mpc\ are connected by edges.}
\label{fig:graph}
\end{figure*}

\begin{figure*}
    \centering
    \includegraphics[width=\linewidth]{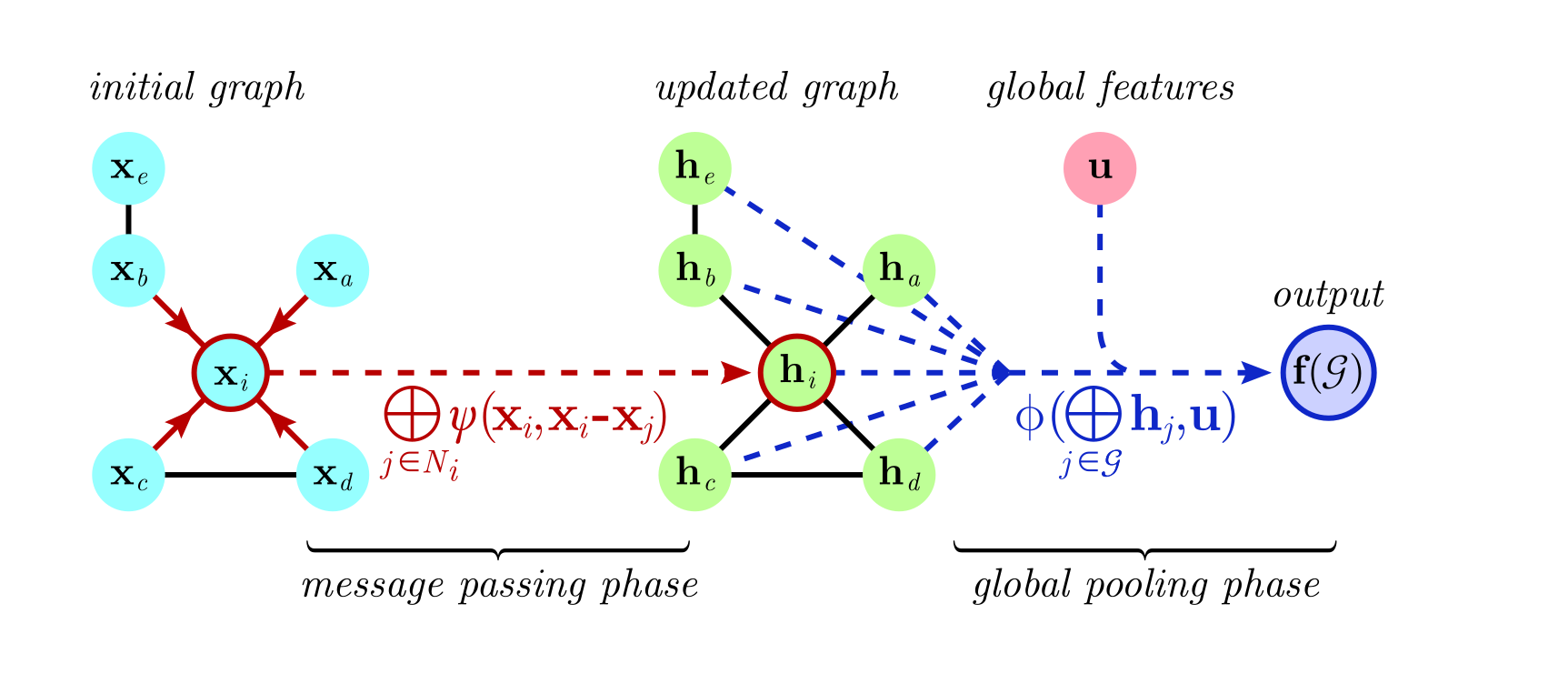}
    \caption{The GNN architecture used in our work. It consists of a message-passing phase in red and a global pooling phase in blue (see Sec. \ref{sec:mygnn} for details). In the message-passing phase, node $i$ receives messages (3D position, stellar mass, star formation rate in our training) $\mathbf{x}$ from its neighbors ($a,b,c,d$) and passes them through an MLP $\psi$. These messages are aggregated to form the hidden feature vector $\mathbf{h}_i$. Next, a global feature $\mathbf{u}$ (number of galaxies in a graph) is added to the hidden feature vector and passed through a second MLP $\phi$. The network outputs the predicted LOS galaxy cluster velocity $\mathbf{f}(\mathcal{G})$. }
    \label{fig:schema_gnn}
\end{figure*}

Fig.~\ref{fig:graph} shows an example of a graph around one cluster. We illustrate the GNN architecture used in our work in Fig.~\ref{fig:schema_gnn}. We adopt edge convolutional layers \citep{Wang2018, Pablo2022}, where the relative vectors $\mathbf{x}_i$-$\mathbf{x}_j$ are combined with the feature vector $\mathbf{x}_i$, replacing Eq.\,\ref{eq:mp} with 
\begin{equation}
    \mathbf{h}_i=\bigoplus_{j\in N_i}\psi(\mathbf{x}_i,\mathbf{x}_i - \mathbf{x}_j). 
\end{equation}
This choice allows us to combine the global shape structure ($\mathbf{x}_i$) with the local  information ($\mathbf{x}_i - \mathbf{x}_j$).
Table~\ref{table:feature} summarizes the features included for each node and in the global-pooling phase. 

\begin{table}
\begin{tabular}{lll} \hline
Type & Feature & Symbol \\ \hline
Node & 3D position & $p$ \\ 
Node & Stellar mass & $m_{*}$ \\ 
Node & Star formation rate & $m_{SFR}$ \\ 
Global & Number of galaxies in a graph & $N_{g}$ \\ \hline
\end{tabular}
\caption{Galaxy properties used in our GNN.}\label{table:feature}
\end{table}

\subsection{GNN Training}
\label{sec:train}

For the training, we first split the Magneticum simulation box of 640 (\mpc)$^3$ into eight independent regions and use seven for training and validation and one for test. The input features such as the 3D positions and galaxy properties (Table~\ref{table:feature}) are provided as part of the simulation. Our network outputs the LOS velocity. We adopt an L2 loss,
\beq
{\it Loss} = \dfrac{1}{N} \sum\limits_{i}^{N} (\varv_{\rm LOS, i}^{\rm truth} - \varv_{\rm LOS, i}^{\rm pred})^2, 
\eeq
where $\varv_{\rm LOS, i}^{\rm truth}$ is the true LOS velocity of {\it i}-th galaxy cluster and $\varv_{\rm LOS, i}^{\rm pred}$ is its predicted value by our GNN.

To optimize the hyperparameters of our network constituting of one aggregation function $\bigoplus$ and two MLPs, $\psi$ and $\phi$, we use the automatic hyperparameter optimization software \textit{Optuna} \citep{Akiba2019}. For the aggregation function $\bigoplus$, we use the mean as it outperforms the sum or the maximum. Our massage-passing MLP ($\phi$) and global-pooling MLP ($\psi$) both consist of three hidden layers with 200, 200, and 100 channels, with ReLU activation functions. The output of the network is the LOS velocity of one galaxy cluster. We adopt the optimization algorithm Adam with a learning rate of $10^{-3}$ and weight decay of $10^{-6}$.


\section{Results}
\label{sec:results}

In this section, we compare the LOS velocities predicted by our GNN to theoretical predictions based on linear perturbation theory and those obtained with CNN in T22. We also investigate the impact of RSD and galaxy number density. 

    \begin{figure*}
    \centering
    \includegraphics[width=0.49\linewidth]{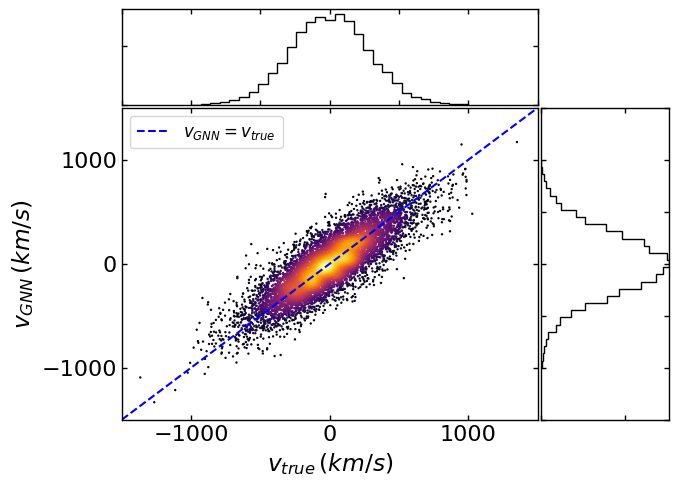}
    \includegraphics[width=0.49\linewidth]{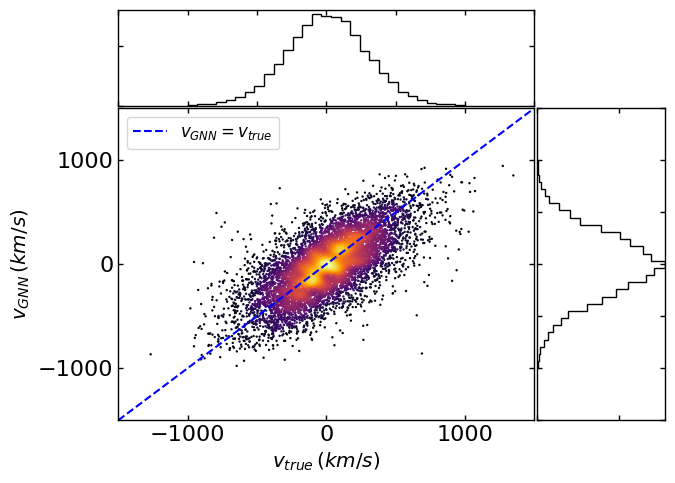}
    \caption{Predicted LOS velocities of mock SDSS clusters by our GNN compared to the true values.
     without RSD (left panel) and with RSD  (right panel). The dashed lines indicate when the two values are equal.} 
    \label{fig:vpred}
    \end{figure*}

\subsection{Reconstructed LOS velocity}
\label{sec:test}

We show in the left panel of Fig.~\ref{fig:vpred} the predicted LOS velocities on the test set ($\approx$12.5\% of the total data), compared with the true velocities. We note that the predicted LOS velocities by our GNN were $\approx$11\% lower than the true values, and this bias is corrected using the procedure in the Appendix. \ref{app:bias}. 
The predicted and true LOS velocities are positively correlated. We quantify the precision of our GNN predictions by 
\beq
\Delta \varv = \sqrt{\frac{1}{N} \sum\limits_{i}^{N} (\varv_{\rm LOS, i}^{\rm truth} - \varv_{\rm LOS, i}^{\rm pred})^2 . }
\eeq
We obtain an uncertainty of $\Delta \varv$=163 km/s. Our GNN outperforms the CNN applied to the SDSS galaxy mocks ($\Delta \varv$=179 km/s, see T22). This improvement is owing to the arbitrarily high spatial resolution that GNN can achieve, compared to the fixed grid size in CNN, and the inclusion of galaxy properties such as the stellar mass (see below). 

\subsection{Comparison with theory}
\label{sec:vel}

To evaluate the performance of our GNN, we compare our results to those from theoretical predictions. The peculiar velocity of a galaxy cluster can be derived in the linear regime by 
\beq
    \bm{\varv}(\bm{k}) = - f(\Omega) a H(a)\,  \frac{i \bm{k}}{k^2} \, \delta(\bm{k}) \, , 
    \label{eq:vel}
\eeq
where $\bm{\varv}(\bm{k})$ and $\delta(\bm{k})$ 
are the velocity and matter density fields in Fourier space, respectively, and $a$ is the scale factor, $H(a)$ is the Hubble parameter, $f(\Omega)$ is the linear velocity growth rate 
given by $f(\Omega) \approx \Omega_{\rm m}^{0.545}$ \citep{Lahav1991, Wang1998, Linder2005, Huterer2007, Ferreira2010}.

We first compute the galaxy density field in 200$^3$~\mpcden\ cubes with a grid size of 5~\mpc\, centered at each cluster. 
We smooth the density field by 15~\mpc\ \citep{Vargas2015} to remove shot noise. We then compute the 3D peculiar velocities of clusters using Eq.\,\ref{eq:vel}. The galaxy bias is found to be $\approx$ 1.3, which is used to translate from the galaxy density field to the matter density field. 
We take the z direction of the 3D velocity in the simulation to be the LOS velocity. The estimated LOS velocities are positively correlated with the true values, with an uncertainty of $\Delta \varv$=181 km/s.

Our GNN again outperforms the linear theory,  as it implicitly models smaller scales and individual galaxy biases without the need to smooth the grid or take an average bias value. 

\subsection{Contribution of galaxy properties}
\label{sec:galprop}

To estimate the contribution from galaxy properties in our velocity reconstruction, we remove galaxy features one by one, such as the total number of galaxies in a graph (within 100~\mpc), the star formation rate, and the stellar mass (Table~\ref{table:feature}). The results are presented in Table~\ref{table:verr}. Adding the stellar mass reduces the velocity uncertainty by 6\%, compared to the position-only GNN. Intuitively, galaxy mass contains information on the gravitational potential, e.g. through the stellar-to-halo mass relation. However, negligible improvements are seen from other features. In our final reported results, we use all the features listed in Table~\ref{table:feature}.

\begin{table}
\begin{tabular}{lcc} \hline
Features & $\Delta \varv$ (km/s)   & $\Delta \varv$  (km/s)\\ 
 & without RSD  & with RSD\\ \hline
$p$ + $m_{*}$ + $m_{SFR}$ + $N_{g}$ & 163 & 210 \\ 
$p$ + $m_{*}$ + $m_{SFR}$ & 164 & 211\\ 
$p$ + $m_{*}$ & 163 & 210\\ 
$p$ & 174 & 217\\ \hline
\end{tabular}
\caption{Uncertainties in GNN predicted cluster LOS velocities.}
\label{table:verr}
\end{table}

\subsection{Effect of redshift-space distortion}
\label{sec:galrsd}

So far, we have input into GNN the galaxy positions in real space. In observations, however, galaxy distances are measured in redshift space. Galaxies' peculiar velocity causes Doppler shifts in addition to the cosmological expansion, resulting in RSD and therefore introducing uncertainties in distance measurement. 

To investigate the effect of RSD in our model, we retrain our network by replacing the comoving positions with redshifts. The redshifts are computed using the LOS velocities of galaxies and clusters. The results are shown in the right panel of Fig.~\ref{fig:vpred}. The predicted LOS velocities using redshifts show a positive correlation with the true LOS velocities but with a larger uncertainty of $\Delta \varv$=210 km/s compared to that without RSD (163 km/s). Even with RSD, our GNN outperforms the CNN results with RSD in T22 ($\Delta \varv$=232 km/s) by $\approx$10\%. 

    \begin{figure}
    \centering
    \includegraphics[width=\linewidth]{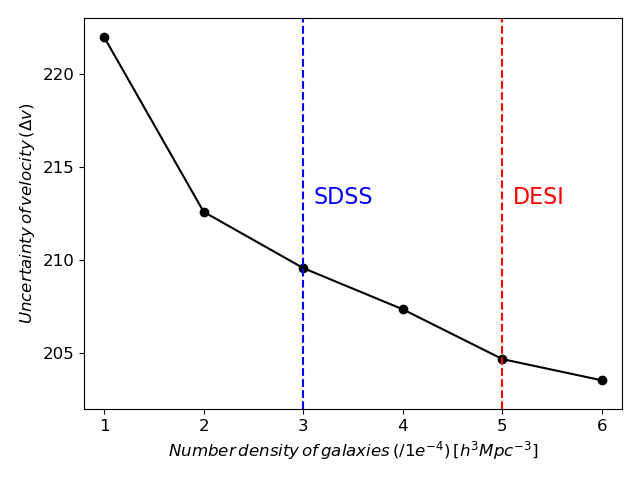}
    \caption{Relative uncertainties of the predicted LOS velocities of clusters by our GNN as a function of galaxy density, with RSD included. The blue and red dashed lines represent the galaxy densities of SDSS and DESI, respectively. } 
    \label{fig:desi}
    \end{figure}

\subsection{Effect of galaxy number density}
\label{sec:ng}

Finally, we investigated the performance of our GNN when the galaxy number density is varied. The number density of SDSS galaxies is about $3 \times 10^{-4}$~\mpcden, or a mean separation distance of $\approx$15~\mpc~\citep{Reid2016}. DESI will see an increase to $\approx$$5 \times 10^{-4}$~\mpcden, or a mean separation distance of $\approx$12.5~\mpc. We expect an improved GNN performance as the galaxy number density increases. We test this assumption with our DESI-like mock (Sec.~\ref{sec:magneticum}). 

We show in Fig.~\ref{fig:desi} the resulting uncertainties in LOS velocity predictions. In addition to the DESI-like galaxies, we vary the galaxy number density from $1 \times10^{-4}$ to $6 \times 10^{-4}$~\mpcden. 
Improvements from increasing the galaxy number density are very mild. For example, for a DESI-like survey, only a $\approx$ 2 \% improvement (from $\Delta \varv$=210 to 205 km/s) is seen, despite an almost 70\% increase in the number density. This shows that using the DESI galaxies instead of SDSS galaxies is unlikely to improve significantly the LOS velocity estimation of clusters.

We compared our results with similar studies done in \cite{Schaan2016, Schaan2021}. They used the Baryon Oscillation Spectroscopic Survey (BOSS) constant-mass (CMASS) galaxies DR10 \citep{Ahn2014} and reconstructed velocities of the galaxies by solving the linearized continuity equation in redshift space \citep{Padmanabhan2012, Vargas2015}.  
The performance of the velocity reconstruction was quantified as a correlation coefficient between the true and reconstructed galaxy velocities, 
\beq 
    r_{\varv} = \frac{\langle \varv_{\rm true} \varv_{\rm rec} \rangle}{\sigma_{\rm \varv}^{\rm true} \sigma_{\rm\varv}^{\rm rec}}, 
\eeq
where $\sigma_{\rm \varv}^{true}$ and $\sigma_{\rm\varv}^{\rm rec}$ are the standard deviations of the true and reconstructed radial velocities, respectively. The value of $r_{\varv}$ was found to be $\approx$0.7 by Refs.~\cite{Schaan2016,Schaan2021} when a Wiener filter was applied. When replaced with a Gaussian filter (as done in \cite{Vargas2015}), \cite{Schaan2021} found that $r_{\varv}$ worsened to $\approx$0.5. 
Our GNN result shows $r_{\varv} \approx$0.74, slightly outperforming these previous works. 
Ref.~\cite{Guachalla2023} performed the velocity reconstruction for the mock DESI galaxies from the AbacusSummit simulation \citep{Maksimova2021} based on the method in \citep{Padmanabhan2012}, predicting $r_{\varv} \approx$0.73. The authors also found little improvement when increasing the galaxy number density beyond that of BOSS (see their Fig.6). Our GNN predicts $r_{\varv} \approx$0.76 for the DESI galaxies, and hence is in line with previous findings. 

\section{Conclusion}
\label{sec:conclusion}

In this work, we investigated possible improvements in reconstructed LOS velocities of clusters from using Graph Neural Networks. In this study, we trained our GNNs to predict the LOS velocities of clusters from the 3D positions and properties of their surrounding galaxies, using realistic SDSS and DESI mock galaxies. 

Our GNNs outperformed both theoretical predictions and CNN by $\approx$10\% (Sec.~\ref{sec:results}), with an uncertainty of $\Delta \varv$=165 km/s. When redshift-space distortion is included, the predictions degrade to $\Delta \varv$=210 km/s , though remain 10\% better than CNN. We found that increasing the number density of galaxies from SDSS to DESI does not significantly improve the LOS velocity reconstruction, unfortunately, in line with previous findings. Therefore, a significant improvement in velocity reconstruction may be more hopeful with improved methods such as the GNNs studied here, rather than a higher galaxy number density. 

GNNs appear to be an exceptionally suitable deep learning architecture to reconstruct the velocity field, owing to several of its advantages:
\begin{itemize}
    \item Galaxy properties can be included in addition to their positions. A 6\%  improvement was seen from adding galaxy masses (Sec.~\ref{sec:galprop}). However, other features (Table.~\ref{table:feature}) did not appear to provide extra information. 
    \item They do not require explicit modeling of the galaxy bias, which is required in traditional methods based on perturbation theory. 
    \item With or without RSD, they outperform CNNs, likely owing to their ability to capture arbitrarily small scales, in comparison to CNN which is limited by the grid size. 
\end{itemize}

In summary, our work demonstrated the potential of GNNs in improving velocity reconstruction with existing and upcoming galaxy observations such as DESI \citep{DESI2016}  and PFS \citep{pfs2016}. Improved velocity measurements will drive our understanding of the gas distribution in our universe when combined with high-precision kSZ measurements, such as those expected from the upcoming CMB experiments Simons Observatory~\citep{Ade2019} and CMB-S4 \citep{Abazajian2019}.

\begin{acknowledgements}

The authors thank Klaus Dolag and Antonio Ragagnin for providing the Magneticum simulations. This work was supported by JSPS KAKENHI Grants 23K13095 and 23H00107 (to JL). 
This work was supported by ILANCE, CNRS - University of Tokyo, International Research Laboratory, Kashiwa, Chiba 277-8582, Japan. This research used computing resources at Kavli IPMU.
SG was partially supported by Beyond AI and ICEPP at University of Tokyo.
\end{acknowledgements}

\bibliographystyle{physrev}
\bibliography{apssamp}

\begin{thebibliography}{10}

\bibitem{Blazek2016}
J.~A. {Blazek}, J.~E. {McEwen}, and C.~M. {Hirata},
\newblock \prl {\bf 116}, 121303 (2016), 1510.03554.

\bibitem{Schmidt2016}
F.~{Schmidt},
\newblock \prd {\bf 94}, 063508 (2016), 1602.09059.

\bibitem{Slepian2018}
Z.~{Slepian} {\em et~al.},
\newblock \mnras {\bf 474}, 2109 (2018), 1607.06098.

\bibitem{Sunyaev1980}
R.~A. {Sunyaev} and I.~B. {Zeldovich},
\newblock \araa {\bf 18}, 537 (1980).

\bibitem{Cooray2001}
A.~{Cooray},
\newblock \prd {\bf 64}, 063514 (2001), astro-ph/0105063.

\bibitem{DeDeo2005}
S.~{DeDeo}, D.~N. {Spergel}, and H.~{Trac},
\newblock arXiv e-prints , astro (2005), astro-ph/0511060.

\bibitem{Shao2011}
J.~{Shao}, P.~{Zhang}, W.~{Lin}, Y.~{Jing}, and J.~{Pan},
\newblock \mnras {\bf 413}, 628 (2011), 1004.1301.

\bibitem{Schaan2016}
E.~{Schaan} {\em et~al.},
\newblock \prd {\bf 93}, 082002 (2016), 1510.06442.

\bibitem{Schaan2021}
E.~{Schaan} {\em et~al.},
\newblock \prd {\bf 103}, 063513 (2021), 2009.05557.

\bibitem{Planck2016IRXXXVII}
{Planck Collaboration},
\newblock \aap {\bf 586}, A140 (2016), 1504.03339.

\bibitem{Munchmeyer2019}
M.~{M{\"u}nchmeyer}, M.~S. {Madhavacheril}, S.~{Ferraro}, M.~C. {Johnson}, and K.~M. {Smith},
\newblock \prd {\bf 100}, 083508 (2019), 1810.13424.

\bibitem{Lim2020}
S.~H. {Lim}, H.~J. {Mo}, H.~{Wang}, and X.~{Yang},
\newblock \apj {\bf 889}, 48 (2020), 1912.10152.

\bibitem{Nguyen2020}
N.-M. {Nguyen}, J.~{Jasche}, G.~{Lavaux}, and F.~{Schmidt},
\newblock \jcap {\bf 2020}, 011 (2020), 2007.13721.

\bibitem{Tanimura2021}
H.~{Tanimura}, S.~{Zaroubi}, and N.~{Aghanim},
\newblock \aap {\bf 645}, A112 (2021), 2007.02952.

\bibitem{Tanimura2022}
H.~{Tanimura}, N.~{Aghanim}, V.~{Bonjean}, and S.~{Zaroubi},
\newblock \aap {\bf 662}, A48 (2022), 2201.01643.

\bibitem{Bolliet2023}
B.~{Bolliet}, J.~{Colin Hill}, S.~{Ferraro}, A.~{Kusiak}, and A.~{Krolewski},
\newblock \jcap {\bf 2023}, 039 (2023), 2208.07847.

\bibitem{Cayuso2023}
J.~{Cayuso}, R.~{Bloch}, S.~C. {Hotinli}, M.~C. {Johnson}, and F.~{McCarthy},
\newblock \jcap {\bf 2023}, 051 (2023), 2111.11526.

\bibitem{Percival2007}
W.~J. {Percival} {\em et~al.},
\newblock \mnras {\bf 381}, 1053 (2007), 0705.3323.

\bibitem{Percival2010}
W.~J. {Percival} {\em et~al.},
\newblock \mnras {\bf 401}, 2148 (2010), 0907.1660.

\bibitem{Beutler2011}
F.~{Beutler} {\em et~al.},
\newblock \mnras {\bf 416}, 3017 (2011), 1106.3366.

\bibitem{Blake2011}
C.~{Blake} {\em et~al.},
\newblock \mnras {\bf 418}, 1707 (2011), 1108.2635.

\bibitem{Anderson2012}
L.~{Anderson} {\em et~al.},
\newblock \mnras {\bf 427}, 3435 (2012), 1203.6594.

\bibitem{Anderson2014}
L.~{Anderson} {\em et~al.},
\newblock \mnras {\bf 441}, 24 (2014), 1312.4877.

\bibitem{Dawson2013}
K.~S. {Dawson} {\em et~al.},
\newblock \aj {\bf 145}, 10 (2013), 1208.0022.

\bibitem{Ross2015}
A.~J. {Ross} {\em et~al.},
\newblock \mnras {\bf 449}, 835 (2015), 1409.3242.

\bibitem{Alam2017}
S.~{Alam} {\em et~al.},
\newblock \mnras {\bf 470}, 2617 (2017), 1607.03155.

\bibitem{Abbott2018}
T.~M.~C. {Abbott} {\em et~al.},
\newblock \mnras {\bf 480}, 3879 (2018), 1711.00403.

\bibitem{Abbott2019}
T.~M.~C. {Abbott} {\em et~al.},
\newblock \mnras {\bf 483}, 4866 (2019), 1712.06209.

\bibitem{Ata2018}
M.~{Ata} {\em et~al.},
\newblock \mnras {\bf 473}, 4773 (2018), 1705.06373.

\bibitem{Bautista2018}
J.~E. {Bautista} {\em et~al.},
\newblock \apj {\bf 863}, 110 (2018), 1712.08064.

\bibitem{Bautista2021}
J.~E. {Bautista} {\em et~al.},
\newblock \mnras {\bf 500}, 736 (2021), 2007.08993.

\bibitem{Agathe2019}
V.~{de Sainte Agathe} {\em et~al.},
\newblock \aap {\bf 629}, A85 (2019), 1904.03400.

\bibitem{Blomqvist2019}
M.~{Blomqvist} {\em et~al.},
\newblock \aap {\bf 629}, A86 (2019), 1904.03430.

\bibitem{Bourboux2020}
H.~{du Mas des Bourboux} {\em et~al.},
\newblock \apj {\bf 901}, 153 (2020), 2007.08995.

\bibitem{Gil2020}
H.~{Gil-Mar{\'\i}n} {\em et~al.},
\newblock \mnras {\bf 498}, 2492 (2020), 2007.08994.

\bibitem{Hou2021}
J.~{Hou} {\em et~al.},
\newblock \mnras {\bf 500}, 1201 (2021), 2007.08998.

\bibitem{Hill2016}
J.~C. {Hill}, S.~{Ferraro}, N.~{Battaglia}, J.~{Liu}, and D.~N. {Spergel},
\newblock \prl {\bf 117}, 051301 (2016), 1603.01608.

\bibitem{Ferraro2016}
S.~{Ferraro}, J.~C. {Hill}, N.~{Battaglia}, J.~{Liu}, and D.~N. {Spergel},
\newblock \prd {\bf 94}, 123526 (2016), 1605.02722.

\bibitem{Munshi2016}
D.~{Munshi}, I.~T. {Iliev}, K.~L. {Dixon}, and P.~{Coles},
\newblock \mnras {\bf 463}, 2425 (2016), 1511.03449.

\bibitem{Park2018}
H.~{Park}, M.~A. {Alvarez}, and J.~R. {Bond},
\newblock \apj {\bf 853}, 121 (2018), 1710.02792.

\bibitem{Kuruvilla2020}
J.~{Kuruvilla}, N.~{Aghanim}, and I.~G. {McCarthy},
\newblock \aap {\bf 644}, A170 (2020), 2010.05911.

\bibitem{Amodeo2021}
S.~{Amodeo} {\em et~al.},
\newblock \prd {\bf 103}, 063514 (2021), 2009.05558.

\bibitem{Lee2022}
B.~K.~K. {Lee}, W.~R. {Coulton}, L.~{Thiele}, and S.~{Ho},
\newblock \mnras {\bf 517}, 420 (2022), 2205.01710.

\bibitem{Zaroubi1995}
S.~{Zaroubi}, Y.~{Hoffman}, K.~B. {Fisher}, and O.~{Lahav},
\newblock \apj {\bf 449}, 446 (1995), astro-ph/9410080.

\bibitem{Zaroubi1996}
S.~{Zaroubi}, A.~{Dekel}, Y.~{Hoffman}, and T.~{Kolatt},
\newblock arXiv e-prints , astro (1996), astro-ph/9603068.

\bibitem{Zaroubi1999}
S.~{Zaroubi}, Y.~{Hoffman}, and A.~{Dekel},
\newblock \apj {\bf 520}, 413 (1999), astro-ph/9810279.

\bibitem{Branchini1999}
E.~{Branchini} {\em et~al.},
\newblock \mnras {\bf 308}, 1 (1999), astro-ph/9901366.

\bibitem{Wang2012}
H.~{Wang}, H.~J. {Mo}, X.~{Yang}, and F.~C. {van den Bosch},
\newblock \mnras {\bf 420}, 1809 (2012), 1108.1008.

\bibitem{Kitaura2012a}
F.-S. {Kitaura}, R.~E. {Angulo}, Y.~{Hoffman}, and S.~{Gottl{\"o}ber},
\newblock \mnras {\bf 425}, 2422 (2012), 1111.6629.

\bibitem{Kitaura2012b}
F.-S. {Kitaura} and R.~E. {Angulo},
\newblock \mnras {\bf 425}, 2443 (2012), 1111.6617.

\bibitem{Benson2000}
A.~J. {Benson}, S.~{Cole}, C.~S. {Frenk}, C.~M. {Baugh}, and C.~G. {Lacey},
\newblock \mnras {\bf 311}, 793 (2000), astro-ph/9903343.

\bibitem{Berlind2002}
A.~A. {Berlind} and D.~H. {Weinberg},
\newblock \apj {\bf 575}, 587 (2002), astro-ph/0109001.

\bibitem{Weinberg2004}
D.~H. {Weinberg}, R.~{Dav{\'e}}, N.~{Katz}, and L.~{Hernquist},
\newblock \apj {\bf 601}, 1 (2004), astro-ph/0212356.

\bibitem{Nuza2013}
S.~E. {Nuza} {\em et~al.},
\newblock \mnras {\bf 432}, 743 (2013), 1202.6057.

\bibitem{Kitaura2016}
F.-S. {Kitaura} {\em et~al.},
\newblock \mnras {\bf 456}, 4156 (2016), 1509.06400.

\bibitem{Rodriguez2016}
S.~A. {Rodr{\'\i}guez-Torres} {\em et~al.},
\newblock \mnras {\bf 460}, 1173 (2016), 1509.06404.

\bibitem{Jackson1972}
J.~C. {Jackson},
\newblock \mnras {\bf 156}, 1P (1972), 0810.3908.

\bibitem{Kaiser1987}
N.~{Kaiser},
\newblock \mnras {\bf 227}, 1 (1987).

\bibitem{Hong2021}
S.~E. {Hong}, D.~{Jeong}, H.~S. {Hwang}, and J.~{Kim},
\newblock \apj {\bf 913}, 76 (2021), 2008.01738.

\bibitem{Wu2021}
Z.~{Wu} {\em et~al.},
\newblock \apj {\bf 913}, 2 (2021), 2105.09450.

\bibitem{Wu2023}
Z.~{Wu} {\em et~al.},
\newblock \mnras {\bf 522}, 4748 (2023), 2301.04586.

\bibitem{Zhou2018}
J.~{Zhou} {\em et~al.},
\newblock arXiv e-prints , arXiv:1812.08434 (2018), 1812.08434.

\bibitem{Pablo2022}
P.~{Villanueva-Domingo} {\em et~al.},
\newblock \apj {\bf 935}, 30 (2022), 2111.08683.

\bibitem{Santi2023}
N.~S.~M. {de Santi} {\em et~al.},
\newblock \apj {\bf 952}, 69 (2023), 2302.14101.

\bibitem{Shao2023}
H.~{Shao} {\em et~al.},
\newblock \apj {\bf 956}, 149 (2023), 2302.14591.

\bibitem{DESI2016}
{DESI Collaboration},
\newblock arXiv e-prints , arXiv:1611.00036 (2016), 1611.00036.

\bibitem{Hirschmann2014}
M.~{Hirschmann} {\em et~al.},
\newblock \mnras {\bf 442}, 2304 (2014), 1308.0333.

\bibitem{Dolag2015}
K.~{Dolag},
\newblock {The Magneticum Simulations, from Galaxies to Galaxy Clusters},
\newblock in {\em IAU General Assembly}, p. 2250156, 2015.

\bibitem{Komatsu2011}
E.~{Komatsu} {\em et~al.},
\newblock \apjs {\bf 192}, 18 (2011), 1001.4538.

\bibitem{Ade2019}
P.~{Ade} {\em et~al.},
\newblock \jcap {\bf 2019}, 056 (2019), 1808.07445.

\bibitem{Abazajian2019}
K.~{Abazajian} {\em et~al.},
\newblock arXiv e-prints , arXiv:1907.04473 (2019), 1907.04473.

\bibitem{Wang2018}
Y.~{Wang} {\em et~al.},
\newblock arXiv e-prints , arXiv:1801.07829 (2018), 1801.07829.

\bibitem{Akiba2019}
T.~{Akiba}, S.~{Sano}, T.~{Yanase}, T.~{Ohta}, and M.~{Koyama},
\newblock arXiv e-prints , arXiv:1907.10902 (2019), 1907.10902.

\bibitem{Lahav1991}
O.~{Lahav}, P.~B. {Lilje}, J.~R. {Primack}, and M.~J. {Rees},
\newblock \mnras {\bf 251}, 128 (1991).

\bibitem{Wang1998}
L.~{Wang} and P.~J. {Steinhardt},
\newblock \apj {\bf 508}, 483 (1998), astro-ph/9804015.

\bibitem{Linder2005}
E.~V. {Linder},
\newblock \prd {\bf 72}, 043529 (2005), astro-ph/0507263.

\bibitem{Huterer2007}
D.~{Huterer} and E.~V. {Linder},
\newblock \prd {\bf 75}, 023519 (2007), astro-ph/0608681.

\bibitem{Ferreira2010}
P.~G. {Ferreira} and C.~{Skordis},
\newblock \prd {\bf 81}, 104020 (2010), 1003.4231.

\bibitem{Vargas2015}
M.~Vargas-Maga\~na, S.~Ho, S.~Fromenteau, and A.~J. Cuesta,
\newblock Mon. Not. Roy. Astron. Soc. {\bf 467}, 2331 (2017), 1509.06384.

\bibitem{Reid2016}
B.~{Reid} {\em et~al.},
\newblock \mnras {\bf 455}, 1553 (2016), 1509.06529.

\bibitem{Ahn2014}
C.~P. {Ahn} {\em et~al.},
\newblock \apjs {\bf 211}, 17 (2014), 1307.7735.

\bibitem{Padmanabhan2012}
N.~{Padmanabhan} {\em et~al.},
\newblock \mnras {\bf 427}, 2132 (2012), 1202.0090.

\bibitem{Guachalla2023}
B.~{Ried Guachalla}, E.~{Schaan}, B.~{Hadzhiyska}, and S.~{Ferraro},
\newblock arXiv e-prints , arXiv:2312.12435 (2023), 2312.12435.

\bibitem{Maksimova2021}
N.~A. {Maksimova} {\em et~al.},
\newblock \mnras {\bf 508}, 4017 (2021), 2110.11398.

\bibitem{pfs2016}
N.~{Tamura} {\em et~al.},
\newblock {Prime Focus Spectrograph (PFS) for the Subaru telescope: overview, recent progress, and future perspectives},
\newblock in {\em Ground-based and Airborne Instrumentation for Astronomy VI}, edited by C.~J. {Evans}, L.~{Simard}, and H.~{Takami}, , Society of Photo-Optical Instrumentation Engineers (SPIE) Conference Series Vol. 9908, p. 99081M, 2016, 1608.01075.

\bibitem{Appleby2023}
S.~{Appleby}, R.~{Dav{\'e}}, D.~{Sorini}, C.~C. {Lovell}, and K.~{Lo},
\newblock \mnras {\bf 525}, 1167 (2023), 2301.02001.

\end{thebibliography}

\appendix

\section{Bias correction}
\label{app:bias}

\begin{figure*}
 \begin{minipage}[c]{0.49\linewidth}
    \centering
\centering
\includegraphics[width=\linewidth]{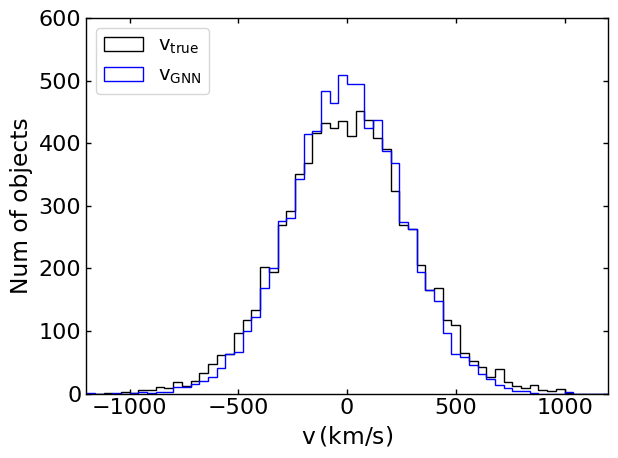}
\includegraphics[width=\linewidth]{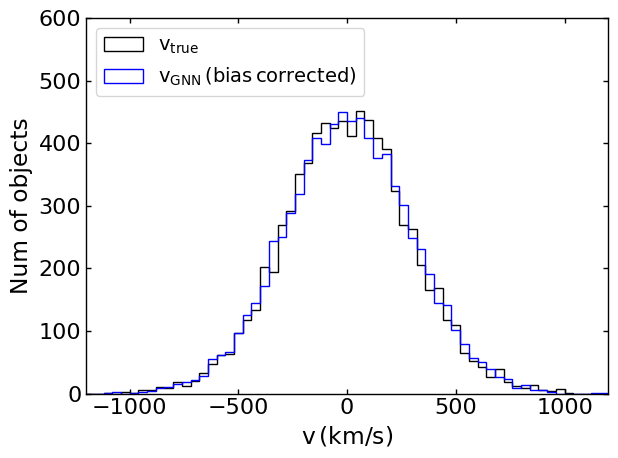}
\caption{Distribution of the true and predicted LOS velocities of clusters before (upper panel) and after (lower panel) the bias correction. }
\label{fig:vbias_hist}
\end{minipage}
 \begin{minipage}[c]{0.49\linewidth}
\centering
\includegraphics[width=\linewidth]{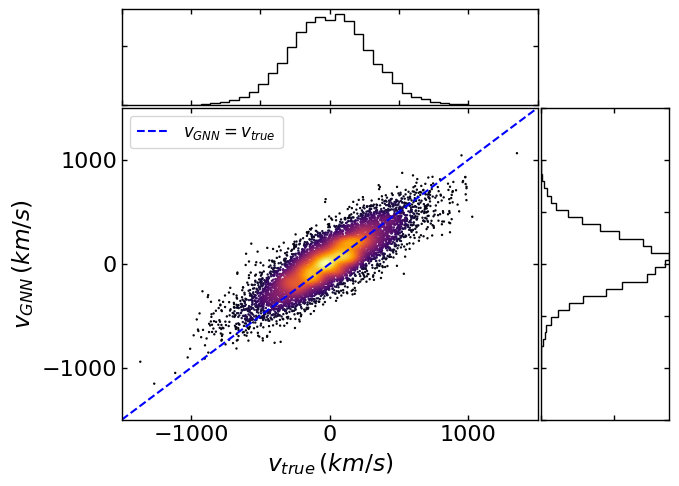}
\includegraphics[width=\linewidth]{outputs_vs_true_biascorr.png}
\caption{Relationship of the true and predicted LOS velocities of clusters before (upper panel) and after (lower panel) the bias correction. } 
\label{fig:vbias_scat}
\end{minipage}
\end{figure*}

We found a positive correlation between the true and predicted values of the LOS velocities of clusters, but the predicted values were slightly biased lower than the true values. We corrected this bias by adding random Gaussian noise to the predictions, as demonstrated by \cite{Appleby2023}. The scale of Gaussian random noise was determined to match the standard deviations of the true and predicted distributions. Fig.~\ref{fig:vbias_hist} shows the distributions of true and predicted values before and after the bias corrections in the left and right panels, respectively. In addition, Fig.~\ref{fig:vbias_scat} shows the scatter plots of the true and predicted values before and after the bias correction in the left and right panels, respectively. These figures show that the bias was statistically corrected to match the true and predicted values. 

\end{document}